\documentstyle[12pt]{article}
\newcommand{\slp}{\raise.15ex\hbox{$/$}\kern-.57em\hbox{$\partial$}}
\newcommand{\sla}{\raise.15ex\hbox{$/$}\kern-.57em\hbox{$a$}}
\newcommand{\slA}{\raise.15ex\hbox{$/$}\kern-.57em\hbox{$A$}}
\newcommand{\slb}{\raise.15ex\hbox{$/$}\kern-.57em\hbox{$b$}}
\newcommand{\be}{\begin{equation}}

\newcommand{\ee}{\end{equation}}
\newcommand{\bear}{\begin{eqnarray}}
\def\eear{\end{eqnarray}}
\newcommand{\ear}{\end{eqnarray}}
\newcommand{\ba}{\begin{eqnarray*}}
\newcommand{\ea}{\end{eqnarray*}}
\begin{document}
\begin{titlepage}
\setcounter{page}{1}
\begin{flushright}
HD--THEP--99--17\\
\end{flushright}
\vskip1.5cm
\begin{center}
{\large{\bf Hamiltonian Approach to Lagrangian Gauge Symmetries}}\\
\vspace{1cm}
R. Banerjee, 
\footnote{email: r.banerjee@thphys.uni-heidelberg.de\\
On leave of absence from S.N. Bose Natl. Ctr. for Basic Sc., Salt Lake, 
Calcutta 700091, India }
H.J. Rothe 
\footnote{email: h.rothe@thphys.uni-heidelberg.de}
and 
K. D. Rothe 
\footnote{email: k.rothe@thphys.uni-heidelberg.de}
\\
{\it Institut  f\"ur Theoretische Physik - Universit\"at Heidelberg}
\\
{\it Philosophenweg 16, D-69120 Heidelberg, Germany}

{(May 3, 1999)}
\end{center}

\begin{abstract}
\noindent
We reconsider the problem of finding all local symmetries of a Lagrangian.
Our approach is completely Hamiltonian without any reference to the 
associated action. We present a simple algorithm for obtaining the
restrictions on the gauge parameters entering in the definition of the
generator of gauge transformations.
\end{abstract}
\end{titlepage}

\newpage

The unravelling of gauge symmetries of a given action is an important problem which has received much attention in the past. Two main approaches have been
followed in the literature: i) the Hamiltonian approach based on Dirac's conjecture \cite{Dirac, Girotti}, where a suitable combination of the first class constraints is shown to be a generator of local symmetries of the
Lagrangian, and ii) a purely Lagrangian approach, based on techniques used for
discussing
differential equations which are unsolvable with respect to the highest
derivatives \cite{Mukunda, Gitman, Chaichian1, Shirzad}.

With regard to the Hamiltonian approach essentially two different procedures
have been followed recently:i) a hybrid approach where one departs from the requirement of the off-shell invariance of the {\it extended} action
\footnote{By extended action we mean the action constructed in terms
of the extended Hamiltonian, in Dirac's terminology.}
under the local symmetry transformations
generated by the phase space constraints, and then imposes a gauge condition whereby all Lagrange multipliers associated with secondary
first-class constraints  vanish \cite{Henneaux}; ii) a purely algebraic approach based on the Poisson algebra of gauge generators with the constraints and the canonical Hamiltonian 
\cite{Gomis,Chaichian2}. In this case the restrictions on the 
gauge parameters have been obtained
only for a special class of constrained systems.

In this paper we present a simple algorithm, based entirely on the {\it total}
Hamiltonian approach, for obtaining the generator of the most general
symmetry transformation of a given action, without ever making an explicit
reference to the action itself. In order to simplify the discussion, and also for the sake of comparison, we restrict 
ourselves in the following to Hamiltonian systems with only
irreducible first class constraints. The extension to systems with mixed first and second
class constraints then involves a trivial step which we shall comment on at the end
of this paper.

Consider a Hamiltonian system of $2n$ degrees of freedom $q_i,p_i, i=1\ldots n$,
described by a canonical Hamiltonian $H_c$ and
a complete and irreducible set of (first class) primary constraints
$\Phi_{a_1} \approx 0\ (a_1 = 1,\cdot\cdot\cdot, r)$,
and secondary constraints $\Phi_{a_2}\approx 0\ (a_2 = r+1,\cdot\cdot\cdot N)$, where $r$ is the rank of the Hessian associated with
the Lagrangian in  question. We collect these constraints into a single
vector with components $\Phi_a, a=1,\cdot\cdot\cdot,N$. Following
Dirac's conjecture, we make the following ansatz for the generator
of gauge transformation:
\be\label{gaugegenerator}
G = \sum_{a=1}^{N} \epsilon^a(t,p,q,v) \Phi_a
\ee
where, as we shall see, it is in general necessary to allow the gauge parameters $\epsilon_a$ to depend
not only explicitely on time, but also
on all phase space variables, including the Lagrange multipliers $\{v^{a_1}\}$
(and their time derivatives) entering
in the total Hamiltonian,
\be\label{totalhamiltonian}
H_T = H_c + \sum_{a_1} v^{a_1}\Phi_{a_1}\;.
\ee
Here $H_c$ is the canonical Hamiltonian, and $\{\Phi_{a_1}\approx 0\}$ are the 
(first class) primary
constraints.

 An infinitesimal gauge transformation of a phase-space function $F(q,p)$ is then given by
\be\label{gaugetransformation}
\delta F = \left[F,\Phi^a\right]\epsilon_a
\ee
where a summation over repeated indices is henceforth implied. Note that the gauge parameters appear outside the Poisson
bracket. 
In principle we could have  included the gauge parameters 
inside the Poisson bracket.  These different
definitions are weakly equivalent. As a result of the algebra
\be
[H_C,\Phi_a] = V_a^b \Phi_b\label{algebra1}
\ee
\be
[\Phi_a,\Phi_b] = C_{ab}^c\Phi_c\label{algebra}
\ee
this weak equivalence
continues to be true for the  Poisson brackets of $\delta F$ with either
the canonical Hamiltonian $H_C$ or the generator $G$. Since our analysis only involves this algebra, it is inconsequential whether the gauge parameters are kept inside or outside the Poisson bracket.  The structure functions
$C_{ab}^c$ and $V_a^b$ can in general depend on the phase space variables.
 
We now notice that the action principle 
which leads to the Euler-Lagrange equations of motion, 
requires the commutativity of a
general $\delta$ variation with the time-differentiation. This commutativity need not hold for an arbitrary variation within the hamiltonian framework. Since our motivation is to abstract the symmetries of the action, we 
impose
\be\label{commutativity1}
\delta \frac{d}{dt}q_i =  \frac{d}{dt}\delta  q_i\ \ \ ; i=1, .... n.
\ee
as a fundamental requirement. This, as we now show, turns out to imply
a non trivial condition on the gauge parameters and Lagrange multipliers.

As shown by Dirac \cite{Dirac}, the Euler-Lagrange equations follow 
from the action principle $\delta S_T=0$, where $S_T$ is defined by
\be\label{action}
S_T = \int dt [p_i\dot q_i - H_T]
\ee
Moreover, the symmetries of the total action $S_T$ are also the symmetries
of  $S=\int dt L(q,\dot q)$, once the Hamilton equation of motion for $\dot q_i$, defining the relation between $\dot q_i$, and the momenta $p_i$ as well as Lagrange multipliers, are used to eliminate the momenta and Lagrange multipliers in favour of all the velocities, including the undetermined ones. 
Since we are interested in the local symmetries of this action, 
we shall thus work with the total Hamiltonian. 

The equations of motion within the hamiltonian framework, are given by,
\be\label{Hamiltonequations}
\dot q_i = [q_i,H_T] = [q_i,H_c] + v^{a_1}[q_i,\Phi_{a_1}]
\ee
and,
\[
\dot p_i = [p_i,H_T] = [p_i,H_c] + v^{a_1}[p_i,\Phi_{a_1}]
\]
together with the constraint equations
\be\label{constraintequations}
\Phi_{a_1} = 0
\ee
 From (\ref{Hamiltonequations}) and (\ref{gaugetransformation}) we obtain for the left hand side of 
(\ref{commutativity1})
\be
\delta \dot q_i = \left[[q_i,H_c],\Phi_a\right]\epsilon^a +
v^{a_1}\left[[q_i,\Phi_{a_1}],\Phi_b\right]\epsilon^b
+ \delta v^{a_1}\left[q_i,\Phi_{a_1}\right]\;\nonumber
\ee
and for the right hand side of the same equation,
\be
\frac{d}{dt}\delta q_i = \left[[q_i,\Phi_a],H_c\right]\epsilon^a
+ v^{a_1}\epsilon^a\left[[q_i,\Phi_a],\Phi_{a_1}\right]
+ \left[q_i,\Phi_a\right]\frac{d\epsilon^a}{dt}\;.
\ee
Equating both expressions and making use of the Jacobi identity,
we obtain
\be\label{relation1}
\left[[H_c,\Phi_a], q_i\right]\epsilon^a 
+ v^{a_1}\left[[\Phi_{a_1},\Phi_{a}],q_i\right]\epsilon^a
- \delta v^{a_1}[q_i,\Phi_{a_1}] + \left[q_i,\Phi_a\right]\frac{d\epsilon^a}{dt}=0\;.
\ee
Using (\ref{algebra1}) and (\ref{algebra}), we see that, on the constraint surface $\{\Phi_a = 0\}$, the above equation implies,

\be\label{relation2}
\left[\frac{d\epsilon^b}{dt}   - \epsilon^a 
[V_a^b  + v^{a_1} C_{a_1a}^b]\right]\frac{\partial\Phi_b}{\partial p_i} - \delta v^{a_1} \frac{\partial\Phi_{a_1}}{\partial p_i}
 = 0\;.\nonumber
\ee

 Now, the first class nature and linear independence (irreducibility) of
the constraints guarantees that each of these can be identified as a momentum conjugate to some coordinate, the precise mapping being effected by a canonical transformation. Since (\ref{relation2}) holds for all $i$ one is led 
to the conditions 
\bear
\delta v^{b_1}&=& \frac{d\epsilon^{b_1}}{dt}   - \epsilon^a 
[V_a^{b_1}  +  v^{a_1} C_{a_1a}^{b_1}]\label{deltalagrangeparameter}\\
 0&=& \frac{d\epsilon^{b_2}}{dt}  - \epsilon^a 
[V_a^{b_2}  +  v^{a_1} C_{a_1a}^{b_2}] \label{conditiononepsilon}
\eear
Note that in the above equations, $\frac{d\epsilon^a}{dt}$ denotes the total time
derivative as given by
\be
\frac{d\epsilon^a}{dt} = \frac{D\epsilon^a}{Dt} + [\epsilon^a,H_T]\;
\ee
where, following the notation of Ref. \cite{Henneaux}, 
\be\label{Dderivative}
\frac{D}{Dt} = \frac{\partial}{\partial t}
+ \dot v^{a_1}\frac{\partial}{\partial v^{a_1}} + 
{\ddot v}^{a_1}\frac{\partial}{\partial {\dot v}^{a_1}} + \cdot\cdot\cdot.
\ee

The restrictions on the gauge parameters and the Lagrange multipliers found here are
seen to agree with that of Ref.\cite{Henneaux},  obtained by looking at the 
invariance of the total action considered as the gauge-fixed version
of the extended action, defined in terms of the extended Hamiltonian
\be\label{extendedhamiltonian}
H_E = H_c + \sum\xi^a \Phi_a\;,
\ee
where the Lagrange multipliers $\{\xi^{a_2}\}$ are required to vanish by imposing suitable gauge conditions.
Our analysis is also equally applicable to a dynamics determined 
by the extended Hamiltonian.
The algebra now involves the full set of (primary and secondary)
first class constraints, so that no condition emerges for the gauge parameters 
while the variation of the Lagrange
multipliers $\xi^a$ is given by
\be
\delta \xi^{a}= \frac{d\epsilon^{a}}{dt}   - \epsilon^b 
[V_b^{a}  +\xi^{c} C_{cb}^{a}]\;.
\ee
Hence one is free to choose the gauge parameters $\epsilon^a$ to be functions of time only.
These equations again agree with those given in reference \cite{Henneaux},
as obtained by requiring the invariance of the corresponding extended action.  

Let us conclude with some comments:
Our requirement (\ref{commutativity1}) only involved 
the relation between the "velocities" and the canonical momenta and the
arbitrary Lagrange multipliers. We have thus only used the "first" of
Hamilton's equations, i.e., (\ref{Hamiltonequations})

Contrary to other procedures, our derivation was carried out on a purely (total) Hamiltonian level.
As such we could have equally well worked with gauge transformations in the form,
\be\label{gaugetransformation2}
\bar\delta F = [F,G] = [F,\epsilon^a\Phi_a]
\ee
because of the first-class nature of $H_c$ and $G$.
However, on the Lagrangian level, the two ways, 
(\ref{gaugetransformation}) and (\ref{gaugetransformation2}) of writing
the gauge transformation matters, since it is to be a symmetry also
away from the constrained surface. As one easily checks, by explicitly looking at the off-shell invariance of the action (\ref{action}), it is the
definition, as given by (\ref{gaugetransformation}),
which leads to the transformation law (\ref{deltalagrangeparameter})
for the Lagrange multipliers, 
and conditions (\ref{conditiononepsilon}) on the gauge parameters. 

Finally we  emphasize that the same discussion applies
to the case where also second class constraints are present, with the
simple replacement of $H_c$ by the first class operator $H^{(1)}$, defined
in the standard way by adding the contribution of the second class
constraints whose Lagrange multipliers are now completely fixed.

\section*{Acknowledgement}

One of the authors (R.B.) would like to thank the Alexander von Humboldt Foundation for providing financial support making this collaboration
possible.


\end{document}